\documentclass[sigconf]{acmart}

\usepackage{amssymb}
\usepackage{amsmath}
\usepackage{mathtools}
\usepackage{bm}
\usepackage{multirow}
\usepackage{longtable}
\usepackage{adjustbox}
\usepackage{enumitem}
\renewcommand{\arraystretch}{1.3}
\usepackage{graphicx}
\usepackage{subcaption}
\usepackage{float}
\usepackage{algorithm}
\usepackage{algpseudocode}
\usepackage{xcolor}
\usepackage[normalem]{ulem}
\useunder{\uline}{\ul}{}
\usepackage{pifont}
\usepackage{ragged2e}
\usepackage{verbatim}
\usepackage{microtype}
\usepackage{setspace}
\usepackage{balance}
\usepackage{url}
\usepackage{hyperref}

\AtBeginDocument{%
  }

\setcopyright{acmlicensed}
\copyrightyear{2018}
\acmYear{2018}
\acmDOI{XXXXXXX.XXXXXXX}
\acmConference[Conference acronym 'XX]{Make sure to enter the correct
  conference title from your rights confirmation email}{June 03--05,
  2018}{Woodstock, NY}
\acmISBN{978-1-4503-XXXX-X/2018/06}

\begin{document}

\title[LoopMemGR]{LoopMemGR: From Behavior Logs to Evolving Memory for Generative Recommendation}

\author{Hui Qian}
\email{yihui.qh@alibaba-inc.com}
\affiliation{%
  \institution{Alibaba Group}
  \city{Beijing}
  \country{China}
}
\authornote{Work done during internship at Alibaba.}

\author{Changfa Wu}
\email{wuchangfa.wcf@alibaba-inc.com}
\affiliation{%
  \institution{Alibaba Group}
  \city{Beijing}
  \country{China}
}
\authornote{Equal contribution.}

\author{Chang Liu}
\email{shimao.lc@alibaba-inc.com}
\affiliation{%
  \institution{Alibaba Group}
  \city{Beijing}
  \country{China}
}

\author{Binbin Cao}
\email{simon.cbb@alibaba-inc.com}
\affiliation{%
  \institution{Alibaba Group}
  \city{Beijing}
  \country{China}
}

\author{Jian Wu}
\email{joshuawu.wujian@alibaba-inc.com}
\affiliation{%
  \institution{Alibaba Group}
  \city{Beijing}
  \country{China}
}

\author{Yuliang Yan}
\email{yuliang.yyl@alibaba-inc.com}
\affiliation{%
  \institution{Alibaba Group}
  \city{Beijing}
  \country{China}
}

\author{Han Zhu}
\email{zhuhan.zh@alibaba-inc.com}
\affiliation{%
  \institution{Alibaba Group}
  \city{Beijing}
  \country{China}
}

\author{Bo Zheng}
\email{bozheng@alibaba-inc.com}
\affiliation{%
  \institution{Alibaba Group}
  \city{Beijing}
  \country{China}
}

\author{Shiye Wang}
\email{wangshiye@ict.ac.cn}
\affiliation{%
  \institution{Institute of Computing Technology, Chinese Academy of Sciences}
  \city{Beijing}
  \country{China}
}
\authornote{Corresponding author.}

\renewcommand{\shortauthors}{Hui Qian et al.}

\begin{abstract}
    Generative recommendation formulates next-item prediction as conditional autoregressive generation over discrete Semantic IDs, enabling end-to-end recommendation over large-scale item spaces. However, most existing methods follow a \emph{history-as-context} paradigm that repeatedly reconstructs user preference from behavior history while discarding system-side recommendation decisions after each request. This creates an \emph{asymmetric memory}: the system remembers what the user has done, but not what it has previously recommended or learned from the resulting feedback. Consequently, useful preference-validation signals, potential negative evidence, and historical exploration information cannot be directly reused across requests.
To address these limitations, we propose \textsc{LoopMemGR}, a closed-loop recommendation experience memory framework for generative recommendation. In addition to the conventional behavior log, \textsc{LoopMemGR} maintains a recommendation experience log that records past recommendation--feedback trajectories. It extracts request-relevant evidence through three complementary views: the \emph{recency view} captures short-term interaction dynamics, the \emph{frequency view} summarizes recurring recommendation patterns, and the \emph{global view} distills transferable regularities shared across users. These signals are compressed into a fixed number of experience tokens to condition the generative backbone under a bounded input budget. Experiments on industrial Taobao and public Amazon datasets demonstrate the effectiveness of closed-loop experience accumulation and multi-view experience extraction.
\end{abstract}

\ccsdesc[500]{Information systems~Recommender systems}

\keywords{Generative Recommendation; Memory-Augmented Recommender Systems; Sequential Recommendation}

\received{20 February 2007}
\received[revised]{12 March 2009}
\received[accepted]{5 June 2009}

\maketitle

\section{Introduction}
\begin{figure}[t]
\centering
\includegraphics[width=\linewidth]{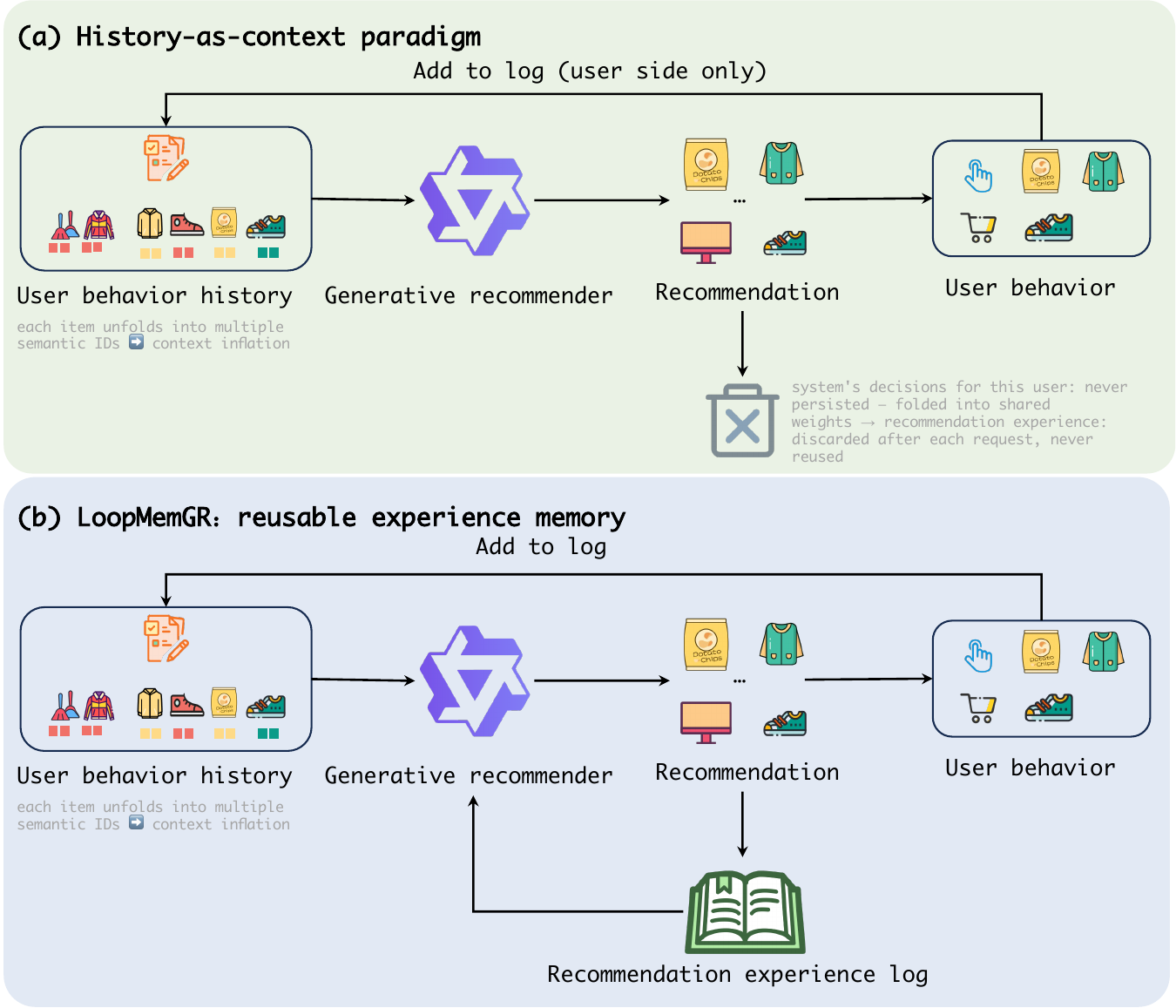}
\caption{Comparison between (a) the prevailing \emph{history-as-context} paradigm and (b) \textsc{LoopMemGR}. In (a), user behaviors are continuously appended to the behavior log, while system-side recommendation decisions are discarded after each request, resulting in an \emph{asymmetric memory}. In (b), \textsc{LoopMemGR} additionally maintains a recommendation experience log, allowing past recommendation--feedback trajectories to be reused in subsequent requests.}
\label{fig:intro}
\end{figure}

Generative recommendation has attracted increasing attention from both academia and industry~\cite{Rajput2023TIGER,Zhai2024HSTU,Wang2024EAGER,Wang2024LETTER}. Unlike discriminative approaches that score items over a predefined candidate set, generative recommenders encode items as discrete Semantic IDs (SIDs) and formulate next-item prediction as conditional autoregressive generation~\cite{Rajput2023TIGER,Wang2024EAGER,Wang2024LETTER}. This paradigm unifies item representation, user modeling, and candidate generation within a single sequence model, enabling end-to-end recommendation over large-scale item spaces~\cite{Rajput2023TIGER,Zhai2024HSTU,Wang2024EAGER,Wang2024LETTER}. 
By directly generating item identifiers, it also provides a flexible interface for integrating heterogeneous user contexts and extending the recommendation vocabulary.

Despite these advances, most existing generative recommenders follow a \emph{history-as-context} paradigm~\cite{Kang2018SASRec,Chang2023TWIN,Lin2024ReLLa,Zhai2024HSTU}. At each request, they reconstruct the user's current preference from historical clicks, purchases, or other behaviors, treating the request as an isolated prediction task. However, real-world recommendation is a continuous interaction process rather than a sequence of independent predictions~\cite{Liu2024DT4IER,Shi2024BiLLP,Zhao2025CaseRec}. Within each interaction round, the system makes recommendation decisions, the user responds to the presented items, and the interaction state evolves accordingly. Although exposure information is sometimes used as auxiliary supervision, the system's previous recommendation decisions are generally not maintained as directly accessible per-user state~\cite{Zhao2025CaseRec,Pan2023SINE,Lv2022XDM}. This creates an \emph{asymmetric memory}: the system remembers what the user has done, but not what it has previously tried.

Beyond behavior history, recommendations made across previous requests form a system-side record of what the recommender has already attempted, which we term \emph{recommendation experience}.Behavior history records realized user actions, whereas recommendation experience records the system-side decision trajectory. The two provide complementary accounts of the interaction process. When considered together, they reveal whether a recommended item subsequently triggered an interaction, whether similar items were repeatedly recommended without response, and which categories or interest regions have already been explored~\cite{Zhao2025CaseRec,Pan2023SINE,Lv2022XDM,Li2023RepExplore,Li2024RightTool}. Without access to such experience, the model must repeatedly reconstruct user preferences from behavior history alone and may unnecessarily repeat previous recommendations or exploration.


A direct solution is to append all historical recommendation results to the generative context. However, recommendation experience grows continuously with the number of requests, while each item may itself consist of multiple SID tokens. Retaining the complete trajectory therefore leads to an unbounded input length and increasing self-attention cost. The key challenge is therefore to summarize the growing recommendation experience and extract request-relevant information under a fixed context budget.

To address this challenge, we propose \textsc{LoopMemGR}, a closed-loop recommendation experience memory framework for generative recommendation. \textsc{LoopMemGR} maintains a recommendation experience log alongside the conventional behavior log: the former stores the system's historical recommendation decisions, while the latter stores realized user interactions. After each request, the newly generated recommendation results are summarized and written into the experience log, and subsequent user responses update the behavior log. The updated information is available only to later requests, preserving causal order and forming a closed loop of recommendation, memory update, and subsequent recommendation.

For each request, \textsc{LoopMemGR} extracts experience memory through three complementary views. The \emph{recency view} retains fine-grained information from recent recommendations to capture the current interaction state. The \emph{frequency view} aggregates recurring recommendation patterns to summarize persistent system judgments and previously explored interests. The \emph{global view} employs learnable queries shared across users to extract transferable regularities when user-specific experience is sparse. The three views are fused into a fixed number of experience tokens, which jointly condition the generative backbone with the behavior history.


The main contributions of this work are summarized as follows:

\begin{itemize}[leftmargin=*]

\item We identify the asymmetric-memory problem in generative recommendation and define the system's historical recommendation-decision trajectory as \emph{recommendation experience}, providing a system-side complement to conventional behavior history.

\item We propose \textsc{LoopMemGR}, which maintains a closed-loop recommendation experience memory and extracts request-relevant information through recency, frequency, and global views under a fixed token budget.

\item We conduct extensive experiments on an industrial Taobao dataset. The results, together with comprehensive ablation and analytical studies, validate the effectiveness of closed-loop experience accumulation, multi-view experience extraction, and fixed-budget experience compression.

\end{itemize}


\section{Method}
In this section, we present \textsc{LoopMemGR} in detail. We first formulate closed-loop generative recommendation with a reusable recommendation experience log, and then describe the overall framework, the unified experience reading operator, the tri-view memory reader and its fusion mechanism, followed by the optimization, causal replay, and complexity analysis.

\subsection{Problem Formulation}

Let \(\mathcal U\) and \(\mathcal V\) denote the user and item sets. For the
\(t\)-th request from user \(u\), the behavior history available before the
request is
\begin{equation}
\mathcal H_u^{<t}=(b_1,b_2,\ldots,b_{L_t}),
\label{eq:history}
\end{equation}
where the behaviors are arranged in chronological order. Following the
generative backbone~\cite{fu2026rankgr}, each item \(v\in\mathcal V\) is
encoded by a two-level Semantic ID from a shared SID codebook,
\(\operatorname{Tok}(v)=\bigl(c^{(1)}(v),c^{(2)}(v)\bigr)\), where the
category-level token \(c^{(1)}\) (SID1) is shared by semantically similar
items and the item-level token \(c^{(2)}\) (SID2) distinguishes items within
the same category. Given the generation context \(\mathbf Z_t\), the
generative recommender \(G_\theta\) predicts the next item \(y_t\) through
two autoregressive steps:
\begin{equation}
p_\theta(y_t\mid\mathbf Z_t)
=p_\theta\!\bigl(c^{(1)}(y_t)\mid\mathbf Z_t\bigr)\,
p_\theta\!\bigl(c^{(2)}(y_t)\mid c^{(1)}(y_t),\mathbf Z_t\bigr).
\label{eq:ar}
\end{equation}

In addition to the conventional behavior log, LoopMemGR maintains a
per-user \emph{recommendation experience log}
\begin{equation}
\mathcal E_u^{<t}=\langle e_1,e_2,\ldots,e_{N_t}\rangle,
\label{eq:exp_log}
\end{equation}
which stores the recommendation experience accumulated before request
\(t\) in exposure order. Each entry corresponds to an item recommended to
\(u\) in an earlier request and is embedded as
\(\mathbf e_j\in\mathbb R^{d}\) using its item representation and log-side
features, yielding \(\mathbf E_t\in\mathbb R^{N_t\times d}\). Because the
experience log grows after every completed request, directly serializing it
into the generative backbone would lead to an unbounded context. LoopMemGR
therefore employs a Tri-View Memory Reader (TVMR) \(f_\psi\) to compress the
log into \(M\) continuous experience tokens,
\(\mathbf T_t=f_\psi(\mathbf E_t)\in\mathbb R^{M\times d}\), where
\(M\ll N_t\). These tokens complement the behavior history in the generation
context, allowing the accumulated recommendation experience to condition
Eq.~\eqref{eq:ar} under a fixed token budget.

\subsection{Framework Overview}

\begin{figure*}[t]
    \centering
    \includegraphics[width=0.9\textwidth,keepaspectratio]{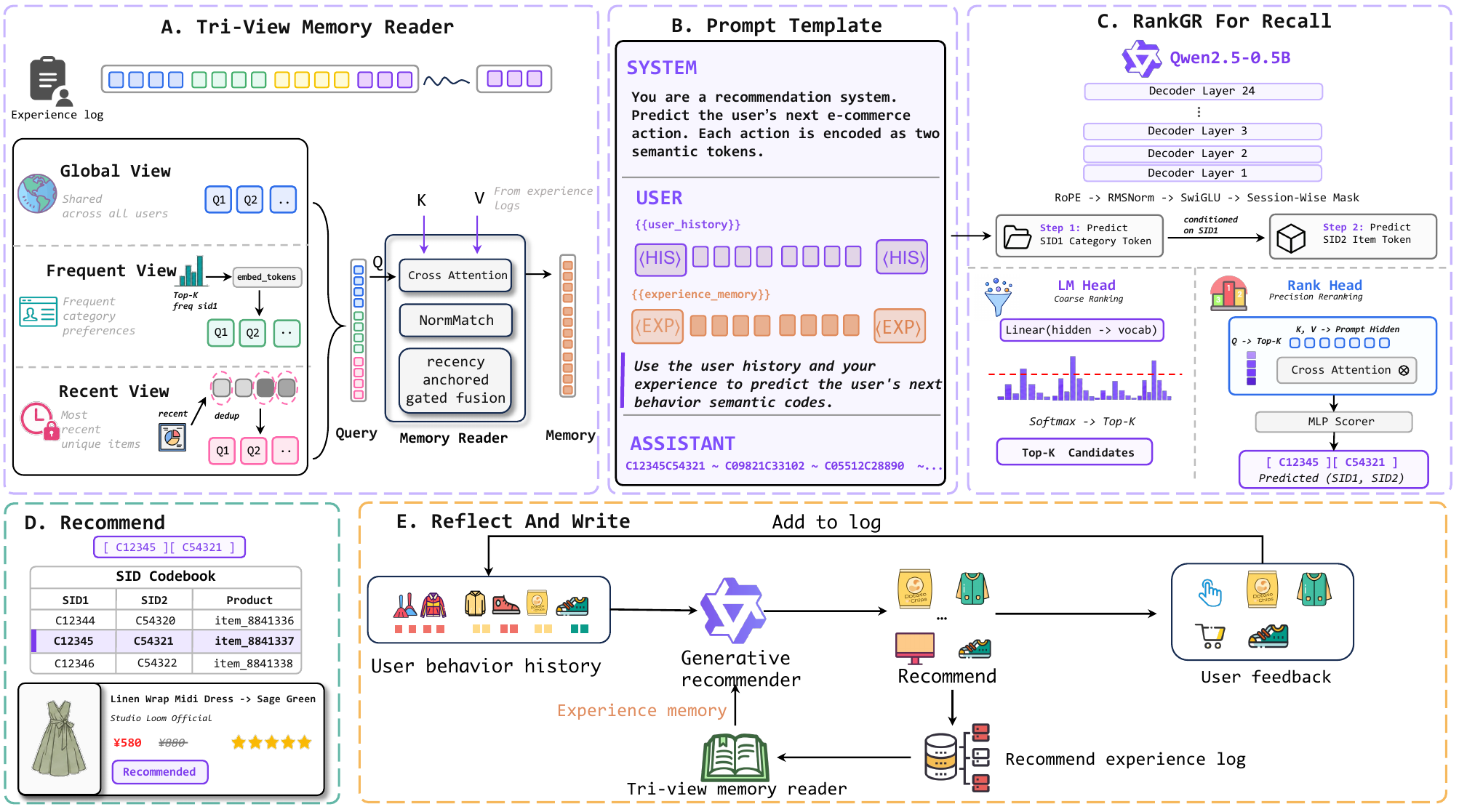}
    \caption{Overview of LoopMemGR. (A)~The Tri-View Memory Reader extracts
    complementary evidence from the accumulated recommendation experience
    log. The recency view preserves fine-grained information from recent
    recommendations, the frequency view captures recurring category-level
    recommendation patterns, and the global view uses learnable queries
    shared across users to extract transferable regularities. Recent-anchored
    gated fusion compresses the three views into a fixed number of experience
    tokens. (B)~The prompt jointly incorporates the user behavior history and
    the experience tokens, with each item represented by two tokens from the
    SID codebook. (C)~The generative backbone predicts the category-level SID1
    followed by the item-level SID2. The LM head retrieves a coarse top-\(K\)
    candidate set, which is reranked by the Rank head through cross-attention to
    the prompt hidden states. (D)~The top-ranked items are presented to the
    user. (E)~After the request, the recommendation results are summarized and
    written into the recommendation experience log, while the subsequent user
    response updates the behavior log, closing the recommendation--feedback--
    memory--recommendation loop.}
    \label{fig:framework}
\end{figure*}

As shown in Figure~\ref{fig:framework}, LoopMemGR processes request \(t\) in
three stages. In \emph{read}, TVMR summarizes the recommendation experience
log \(\mathcal E_u^{<t}\) into \(M\) experience tokens \(\mathbf T_t\)
through the recency, frequency, and global views. In \emph{recommend}, the
behavior history and experience tokens are serialized into \(\mathbf Z_t\),
from which \(G_\theta\) generates and ranks a candidate set \(\mathcal C_t\).
In \emph{memory update}, a frozen summarization module \(S_\phi\) converts
the current recommendation decisions into an ordered item summary \(s_t\),
which is appended to \(\mathcal E_u\), while the user's subsequent feedback
is incorporated into \(\mathcal H_u\) through the standard pipeline. Both
logs are updated only after request \(t\); consequently, the resulting
recommendation--feedback trajectory becomes available from request \(t+1\)
onward and cannot leak into the request that produced it.

\subsection{Unified Experience Reading Operator}
\label{sec:method_reader}

All three views instantiate a shared reading operator that extracts evidence
from a potentially long recommendation experience log. The operator combines
cosine cross-attention with a gated, magnitude-bounded residual update. Let
\(\mathbf B=[\mathbf b_1;\ldots;\mathbf b_q]\in\mathbb R^{q\times d}\)
denote a set of base queries and let
\(\mathbf H=[\mathbf h_1;\ldots;\mathbf h_n]\in\mathbb R^{n\times d}\)
denote the log entries to be read. For each query and key, we first apply RMS
normalization and learned projections, followed by row-wise \(\ell_2\)
normalization:
\begin{equation}
\widehat{\mathbf q}_i=
\frac{W_Q\operatorname{RMSNorm}(\mathbf b_i)}
{\|W_Q\operatorname{RMSNorm}(\mathbf b_i)\|_2},\qquad
\widehat{\mathbf k}_j=
\frac{W_K\operatorname{RMSNorm}(\mathbf h_j)}
{\|W_K\operatorname{RMSNorm}(\mathbf h_j)\|_2}.
\label{eq:qk_normalization}
\end{equation}
The attention scores and normalized weights are
\begin{equation}
s_{ij}=\frac{\widehat{\mathbf q}_i^\top\widehat{\mathbf k}_j}{\tau}
+p_{ij}+m_j,
\qquad
a_{ij}=\frac{\exp(s_{ij})}{\sum_{k=1}^{n}\exp(s_{ik})},
\label{eq:cosine_attention}
\end{equation}
where \(\tau\) is a constrained learnable temperature, \(p_{ij}\) is an
optional view-specific attention prior, and \(m_j\) is zero for a valid log
position and \(-\infty\) for padding. The content update for query \(i\) is
\begin{equation}
\boldsymbol\delta_i=W_O\sum_{j=1}^{n}a_{ij}W_V\mathbf h_j,
\qquad
\Delta(\mathbf B,\mathbf H;\mathbf P)
=[\boldsymbol\delta_1;\ldots;\boldsymbol\delta_q].
\label{eq:attention_update}
\end{equation}
The normalized queries and keys bound the similarity scale, while the value
projections preserve the content carried by the original experience
representations.

Because the length and quality of an experience log may vary substantially,
an unrestricted residual update can overwhelm the base queries. We therefore
define the matrix RMS and a magnitude cap as
\begin{equation}
\operatorname{RMS}(\mathbf Y)
=\sqrt{\frac{\|\mathbf Y\|_F^2}{|\mathbf Y|}},\qquad
\operatorname{Cap}(\Delta;\mathbf B,\rho)
=\Delta\min\!\left(1,
\frac{\rho\operatorname{RMS}(\mathbf B)}
{\operatorname{RMS}(\Delta)+\epsilon}\right),
\label{eq:cap}
\end{equation}
where \(|\mathbf Y|\) is the number of scalar elements in \(\mathbf Y\),
\(\rho\) bounds the update magnitude relative to the base, and
\(\epsilon>0\) is a numerical stabilizer. We further use
\begin{equation}
\operatorname{NormMatch}(\mathbf Y;\mathbf B)
=\mathbf Y\frac{\operatorname{RMS}(\mathbf B)}
{\operatorname{RMS}(\mathbf Y)+\epsilon}
\label{eq:norm_match}
\end{equation}
to restore the RMS scale of the base without altering the direction of the
updated representation. The shared reading operator is therefore
\begin{equation}
\operatorname{Read}(\mathbf B,\mathbf H;\mathbf P)
=\operatorname{NormMatch}\!\left(
\mathbf B+\sigma(g)\operatorname{Cap}\!\left(
\Delta(\mathbf B,\mathbf H;\mathbf P);\mathbf B,\rho\right);\mathbf B
\right),
\label{eq:read}
\end{equation}
where \(g\) is a learnable gate logit. The gate controls whether the retrieved
content should be incorporated, while the cap limits its maximum influence.
Both operations can attenuate an update but do not amplify an already small
one. We omit \(\mathbf P\) when no view-specific prior is used.

\subsection{Tri-View Memory Reader}
\label{sec:method_tvmr}

Recommendation experience contains complementary evidence at different
levels. Recent recommendations reflect short-term interests and the current
interaction context; patterns recurring across requests summarize more
persistent system judgments; and regularities shared across users provide
transferable population-level priors. TVMR accordingly reads the experience
log through recency, frequency, and global views. Each view produces at most
\(M\) tokens, which are subsequently fused into the same fixed \(M\)-token
budget.

\paragraph{Recency view.}
The recency view preserves fine-grained information from the most recent
recommendations. Because repeated exposures of the same item are common,
using the raw tail of the log may allocate multiple memory positions to the
same item. We therefore deduplicate the log while retaining the latest
occurrence of each distinct item. Let
\(\operatorname{Uniq}(\mathbf E_t)\) denote the resulting sequence. For
\(|\operatorname{Uniq}(\mathbf E_t)|>M\), we split it into the latest \(M\)
unique entries and the earlier remainder:
\begin{equation}
\mathbf R_0=\operatorname{Tail}_M\!\bigl(\operatorname{Uniq}(\mathbf E_t)\bigr),
\qquad
\mathbf E_{\mathrm{old}}
=\operatorname{Uniq}(\mathbf E_t)\setminus\mathbf R_0.
\label{eq:recent_split}
\end{equation}
The recent entries preserve their exposure order and serve as the memory
anchor. Using them as queries, the recency view reads the earlier experience:
\begin{equation}
\mathbf R=\operatorname{Read}(\mathbf R_0,\mathbf E_{\mathrm{old}}),
\label{eq:recent_view}
\end{equation}
which prevents a query from directly retrieving its identical occurrence
from the key--value sequence. When the deduplicated log contains at most
\(M\) entries, TVMR retains all of them directly rather than introducing
additional memory tokens.

\paragraph{Frequency view.}
The frequency view captures category-level recommendation patterns that
recur across requests. Let \(c(e_j)=c^{(1)}(e_j)\) be the category-level
identifier (SID1) of the recommended item in entry \(e_j\). For every
category \(c\) in the log, let \(n_c\) denote its occurrence count and
\(p_c\) its latest occurrence position. Categories are ordered
lexicographically by
\begin{equation}
\pi(c)=\bigl(\mathbb{I}[n_c>1],\,n_c,\,p_c\bigr),
\label{eq:frequency_priority}
\end{equation}
which prioritizes categories recommended repeatedly across requests and
breaks ties by frequency and recency. The top \(M\) categories form the
query set \(\mathcal C_F\). For the \(i\)-th selected category \(c_i\), the
initial query is
\begin{equation}
\mathbf f_i^{(0)}=
\mathbf e_{\mathrm{sid}}(c_i)
+\mathbf e_{\mathrm{freq}}(\operatorname{bucket}(n_{c_i}))
+\eta\mathbf e_{\mathrm{type}}(F).
\label{eq:frequency_query}
\end{equation}
Here \(F\) identifies the frequency view, \(\eta\) controls the contribution
of its source embedding, and \(\mathbb{I}[\cdot]\) denotes the indicator
function. A nonnegative learnable bias encourages each query to retrieve
experience entries from the corresponding category:
\begin{equation}
p_{ij}^{F}=\beta\mathbb{I}[c_i=c(e_j)],
\qquad \beta=\operatorname{softplus}(\widetilde\beta),
\label{eq:frequency_bias}
\end{equation}
and the frequency representation is
\begin{equation}
\mathbf F=\operatorname{Read}(\mathbf F^{(0)},\mathbf E_t;\mathbf P^F).
\label{eq:frequency_view}
\end{equation}
Because Eq.~\eqref{eq:frequency_bias} introduces a soft prior rather than a
hard mask, a query may still attend to other categories when supported by
the task objective. The resulting representation summarizes recurring
category-level evidence associated with the system's more persistent
judgments about the user.

\paragraph{Global view.}
User-specific recommendation experience can be sparse, particularly during
the early stages of interaction. The global view therefore introduces \(M\)
learnable queries shared across all users,
\begin{equation}
\mathbf G^{(0)}=[\mathbf g_1^{(0)};\ldots;\mathbf g_M^{(0)}],
\label{eq:global_queries}
\end{equation}
which are initialized with separated query directions and optimized over the
experience logs of the entire user population. Two consecutive reads allow
the shared queries to extract transferable recommendation regularities from
the full user-specific log:
\begin{equation}
\mathbf G^{(1)}=\operatorname{Read}(\mathbf G^{(0)},\mathbf E_t),
\qquad
\mathbf G=\operatorname{Read}(\mathbf G^{(1)},\mathbf E_t).
\label{eq:global_view}
\end{equation}
The resulting tokens are latent, task-oriented representations and are not
assumed to have a one-to-one correspondence with human-interpretable
patterns.

\subsection{Recent-Anchored Gated Fusion}

The three views jointly produce \(3M\) candidate tokens, whereas the
generative backbone accepts only \(M\) experience tokens. To preserve a
bounded working memory, we use the recent entries \(\mathbf R_0\) as the
output anchors and integrate complementary evidence from all three views.
After adding a source embedding to each view, \(\mathbf R_0\) reads the three
branches separately:
\begin{equation}
\Delta_R=\operatorname{Attn}(\mathbf R_0,\widetilde{\mathbf R}),\quad
\Delta_F=\operatorname{Attn}(\mathbf R_0,\widetilde{\mathbf F}),\quad
\Delta_G=\operatorname{Attn}(\mathbf R_0,\widetilde{\mathbf G}).
\label{eq:view_updates}
\end{equation}
Here \(\operatorname{Attn}\) denotes Eqs.~\eqref{eq:qk_normalization}--
\eqref{eq:attention_update} without the residual update. For the recency
view, diagonal query--key pairs are masked so that the \(i\)-th output
position cannot directly copy the \(i\)-th recent candidate.

Each view has an independent gate, and its update is capped before the three
sources are combined:
\begin{equation}
\Delta_{\mathrm{mix}}=
\sum_{v\in\{R,F,G\}}\alpha_v
\operatorname{Cap}(\Delta_v;\mathbf R_0,\rho_s),
\qquad \alpha_v=\sigma(g_v).
\label{eq:mixed_update}
\end{equation}
The combined correction is capped again at the output level:
\begin{equation}
\mathbf T_t=\operatorname{NormMatch}\!\left(
\mathbf R_0+
\operatorname{Cap}(\Delta_{\mathrm{mix}};\mathbf R_0,\rho_m);
\mathbf R_0\right).
\label{eq:fuse}
\end{equation}
Accordingly, TVMR represents the frequency and global evidence as bounded
corrections to a recent-experience anchor, rather than reconstructing the
working memory entirely from unconstrained learned queries. When either view
provides little request-relevant evidence, its gate can suppress the
corresponding update, keeping Eq.~\eqref{eq:fuse} close to \(\mathbf R_0\).

\subsection{Training Protocol and Complexity Analysis}

TVMR and the generative backbone are optimized jointly. Following the
backbone training setting~\cite{fu2026rankgr}, the task objective consists of
the generation loss over the two SID steps and the ranking loss of the Rank
head:
\begin{equation}
\mathcal L_{\mathrm{task}}
=\mathcal L_{\mathrm{gen}}+\mathcal L_{\mathrm{rank}}.
\label{eq:task_loss}
\end{equation}
The experience summarization module remains frozen, and writing the resulting
summary to the recommendation experience log is a discrete state update through
which no gradient is propagated.

To discourage collapse among the global queries and their corresponding
attention distributions, let \(\mathbf g_i\) denote the \(i\)-th global
query and \(\mathbf a_i\) its attention distribution over the recommendation
experience log. We introduce
\begin{align}
\mathcal L_Q
&=\frac{1}{M(M-1)}\sum_{i\ne j}
\left(\frac{\mathbf g_i^\top\mathbf g_j}
{\|\mathbf g_i\|_2\|\mathbf g_j\|_2}\right)^2, \\
\mathcal L_A
&=\frac{1}{M(M-1)}\sum_{i\ne j}
\left(\frac{\mathbf a_i^\top\mathbf a_j}
{\|\mathbf a_i\|_2\|\mathbf a_j\|_2}\right)^2.
\label{eq:diversity_losses}
\end{align}
The final objective is
\begin{equation}
\mathcal L=\mathcal L_{\mathrm{task}}
+\lambda_Q w(s)\mathcal L_Q
+\lambda_A w(s)\mathcal L_A,
\label{eq:loss}
\end{equation}
where \(w(s)\) gradually introduces the diversity regularizers during
training, and \(\lambda_Q\) and \(\lambda_A\) control their respective
strengths.

Training and offline replay follow the same request chronology as online
serving. For request \(t\), the model can access only
\(\mathcal H_u^{<t}\) and \(\mathcal E_u^{<t}\). The current user feedback
and recommendation summary are written only after candidate generation.
This causal ordering prevents future behaviors or recommendation decisions
from entering the context of an earlier request.

For an experience log of length \(N_t\), one cross-attention read costs
\(O((N_t+M)d^2+M N_t d)\). TVMR invokes a fixed number of such reads and does
not perform quadratic self-attention over the full experience log. The
experience component of the backbone context always occupies \(M\) tokens
plus \(O(1)\) markers, independent of the number of completed requests,
while the behavior history follows the backbone's standard truncation
strategy. Appending one recommendation summary costs \(O(m)\). Although the
reader still scans the full log linearly, the fixed experience-token budget
bounds the context processed by the deep generative backbone.

\section{Experiment}
In this section, we conduct a series of experiments to answer the
following research questions:

\noindent\textbf{RQ1: Overall effectiveness.}
How does LoopMemGR perform compared with representative conventional
and generative retrieval methods?

\noindent\textbf{RQ2: Closed-loop experience memory.}
Does persisting recommendation experience across requests improve
subsequent recommendation, and can TVMR retain this benefit within a
fixed 16-token budget?

\noindent\textbf{RQ3: Query division in the global view.}
Do the diversity regularizers prevent the global queries from
collapsing onto the same evidence?

\noindent\textbf{RQ4: View-level reading patterns.}
Do the three views extract complementary evidence from the
experience log, as designed?

\subsection{Experimental Setup}
\label{sec:experimental_setup}

\subsubsection{Datasets.}

For brevity, details of the datasets are deferred to Section~\ref{dataset}.

\subsubsection{Baselines.}
For brevity, details of the baselines are deferred to Section~\ref{baselines}.

\subsubsection{Evaluation Metrics.}

Following RankGR~\cite{fu2026rankgr}, we use Hit Rate at cutoff $K$
(HR@$K$) as the primary retrieval metric. Given the top-$K$ retrieved
item set $\mathcal{I}_u^K$ and the ground-truth interaction set
$\mathcal{I}_u^{\mathrm{gt}}$ for user $u$, HR@$K$ is defined as

\begin{equation}
\mathrm{HR}@K
=
\frac{1}{|\mathcal{U}|}
\sum_{u\in\mathcal{U}}
\frac{
    \left|
        \mathcal{I}_u^K
        \cap
        \mathcal{I}_u^{\mathrm{gt}}
    \right|
}{
    \left|
        \mathcal{I}_u^{\mathrm{gt}}
    \right|
},
\label{eq:hit_rate}
\end{equation}

where $\mathcal{U}$ denotes the evaluated user set. Higher HR indicates
that a larger proportion of ground-truth items is covered by the
retrieved candidate set.

On the \emph{Taobao} dataset, the overall comparison evaluates two
behavior-level targets: click and page view (PV). We report
$\mathrm{HR}^{\mathrm{Click}}@K$ and
$\mathrm{HR}^{\mathrm{PV}}@K$ at
$K\in\{20,100,500,1000,2000\}$. In the experience-memory analysis,
we report the same click and PV targets at
$K\in\{20,1000\}$ using the same HR definition. All HR values are
presented as percentages. Methods appearing in the same table are
evaluated with the same target definitions, candidate corpus, and
retrieval cutoffs.

\subsubsection{Implementation Details.}

For all baseline methods, we follow the implementations, data
partitions, and hyperparameter settings used by
RankGR~\cite{fu2026rankgr}. LoopMemGR uses RankGR as its generative
recommendation backbone. In particular, semantic identifiers are
constructed following FORGE~\cite{FORGE}, and the generative model is
initialized from Qwen2.5-0.5B-Instruct~\cite{qwen2024qwen25}. Each item
is represented by a two-level semantic identifier.

Following RankGR, the raw behavioral input contains at most 2,000 SID
tokens, corresponding to a maximum of 1,000 historical items. The
maximum training sequence length is 4,352 tokens, with the source and
target budgets set to 4,096 and 256 tokens, respectively. LoopMemGR
keeps this behavioral input unchanged and additionally reads the
experience log through TVMR, which outputs $M=16$ experience tokens.
The hidden dimension of all memory representations is aligned with
that of the RankGR backbone.

The experience log is append-only. After request $t$ is completed,
the reflected item sequence is appended to the log in its emitted
order, and repeated items are retained as separate entries. The write
operation uses no learned importance score, temporal decay, or
memory-side re-ranking. At reading time, TVMR considers up to the
most recent 1,000 experience entries and always returns 16 experience
tokens. Moreover, the experience written at request $t$ is visible
only to requests from $t+1$ onward.

We train the models for one epoch on \emph{Taobao}.
The per-device batch size is set to 40. We use a linear
learning-rate schedule with an initial learning rate of
$5\times10^{-5}$ and conduct training in bfloat16 precision. During
inference, dynamic beam search is applied to the two SID levels, with
beam sizes of 500 and 1,400, respectively. The candidate-retention
sizes in the refined scoring phase of RankGR are set to
$\{1400,1400\}$. Unless otherwise stated, the same LoopMemGR
configuration is used in all experiments.

\subsection{Overall Performance (RQ1)}
\label{sec:overall_performance}

We compare LoopMemGR with conventional and generative retrieval
methods on two datasets: the industrial \emph{Taobao} dataset and
the public \emph{Beauty} benchmark. The results are reported in
Tables~\ref{tab:taobao_overall} and~\ref{tab:beauty_overall}, where
the best and second-best results in each column are marked in bold
and underlined, respectively. We make the following observations:

\begin{table*}[t]
\caption{
Overall performance on the industrial \emph{Taobao} dataset.
Baseline results are quoted from~\cite{fu2026rankgr} under the same
evaluation protocol. All values are reported as percentages.
}
\label{tab:taobao_overall}
\centering
\small
\setlength{\tabcolsep}{5.5pt}
\begin{tabular}{l ccccc c ccccc}
\toprule
& \multicolumn{5}{c}{Click HR@$K$}
& &
\multicolumn{5}{c}{PV HR@$K$} \\
\cmidrule{2-6}
\cmidrule{8-12}
Method
& $K{=}20$
& $K{=}100$
& $K{=}500$
& $K{=}1000$
& $K{=}2000$
&
& $K{=}20$
& $K{=}100$
& $K{=}500$
& $K{=}1000$
& $K{=}2000$ \\
\midrule

YouTubeDNN
& 2.33
& 5.06
& 9.05
& 10.80
& 11.86
&
& 0.29
& 2.01
& 5.61
& 7.17
& 7.43 \\

SASRec
& 3.53
& 7.88
& 12.68
& 15.42
& 17.08
&
& 0.42
& 2.13
& 7.87
& 13.12
& 16.27 \\

BERT4Rec
& 3.54
& 7.91
& 12.79
& 15.50
& 17.24
&
& 0.32
& 2.25
& 8.43
& 13.72
& 16.62 \\

Caser
& 3.00
& 7.52
& 11.50
& 14.39
& 15.86
&
& 0.30
& 1.99
& 7.40
& 12.42
& 15.30 \\

NextItNet
& 3.39
& 7.71
& 12.32
& 15.15
& 16.74
&
& 0.33
& 2.08
& 7.53
& 12.59
& 15.67 \\

CORE
& 3.98
& 8.67
& 15.50
& 18.50
& 20.29
&
& 0.41
& 2.47
& 8.59
& 13.97
& 17.26 \\

HSTU
& 4.06
& 8.74
& 15.70
& 18.64
& 20.48
&
& 0.59
& 2.78
& 9.58
& 15.08
& 19.51 \\

TIGER
& 7.17
& 15.59
& 27.89
& 33.27
& 36.53
&
& 1.89
& 6.75
& 18.92
& 25.29
& 31.06 \\

FORGE
& 9.04
& 19.44
& 36.05
& 43.09
& 47.15
&
& 2.47
& 8.39
& 21.90
& 29.74
& 35.37 \\

RankGR
& \underline{11.64}
& \underline{25.30}
& \underline{45.25}
& \underline{54.00}
& \underline{59.28}
&
& \underline{2.99}
& \underline{10.66}
& \underline{28.13}
& \underline{38.18}
& \underline{45.46} \\

\midrule

\textbf{LoopMemGR}
& \textbf{20.43}
& \textbf{39.24}
& \textbf{60.55}
& \textbf{68.41}
& \textbf{70.85}
&
& \textbf{7.16}
& \textbf{22.53}
& \textbf{45.78}
& \textbf{53.56}
& \textbf{62.18} \\

\bottomrule
\end{tabular}
\end{table*}

\begin{table}[t]
\caption{
Overall performance on the public \emph{Beauty} benchmark.
Baseline results are obtained under the same evaluation protocol.
All values are reported as percentages.
}
\label{tab:beauty_overall}
\centering
\small
\setlength{\tabcolsep}{5.5pt}
\renewcommand{\arraystretch}{1.05}
\resizebox{\columnwidth}{!}{%
\begin{tabular}{l cccc}
\toprule
Method
& HR@$20$
& HR@$50$
& HR@$100$
& HR@$500$ \\
\midrule
YouTubeDNN & 3.57 & 6.71 & 9.19 & 19.67 \\
SASRec     & 3.72 & 7.06 & 9.76 & 21.19 \\
BERT4Rec   & 3.70 & 7.11 & 9.79 & 21.21 \\
Caser      & 3.55 & 6.64 & 9.09 & 19.53 \\
NextItNet  & 3.51 & 6.61 & 9.02 & 19.56 \\
CORE       & 3.02 & 6.69 & 9.19 & 21.35 \\
HSTU       & 3.74 & 7.14 & 9.93 & 21.67 \\
TIGER      & 2.09 & 4.96 & 5.90 & 13.12 \\
FORGE      & 4.35 & 7.61 & 10.88 & 22.50 \\
RankGR
& \underline{4.66}
& \underline{7.84}
& \underline{11.17}
& \underline{22.89} \\
\midrule
\textbf{LoopMemGR}
& \textbf{6.79}
& \textbf{11.47}
& \textbf{15.95}
& \textbf{30.69} \\
\bottomrule
\end{tabular}%
}
\end{table}

\begin{itemize}[leftmargin=*, itemsep=2pt, topsep=3pt]
    \item \textbf{LoopMemGR achieves the best performance across
    all evaluation settings.}
    It consistently ranks first for both click and PV prediction at
    every cutoff. Compared with RankGR, LoopMemGR improves Click
    HR by 11.57--15.30 percentage points at $K\geq100$ and improves
    PV HR by 11.87--17.65 percentage points over the same range.
    These results show that the proposed memory framework benefits
    different feedback targets and retrieval depths.

    \item \textbf{The advantage of LoopMemGR transfers to the public
    benchmark.}
    On \emph{Beauty}, where the experience memory is reconstructed
    via cross-fold exposure replay (Appendix~\ref{app:public_replay}),
    LoopMemGR again achieves the best performance, improving HR@20
    from 4.66 to 6.79 over RankGR. The gain is obtained without any
    production exposure log, indicating that the benefit comes from
    closing the loop over the system's own decisions rather than
    from dataset-specific signals.

    \item \textbf{LoopMemGR consistently improves upon its
    backbone.}
    RankGR is the strongest baseline and shares the semantic
    identifiers, generative backbone, training objective, and
    decoding pipeline with LoopMemGR. The consistent improvements
    over RankGR therefore support attributing the performance gains
    to the added behavioral intent and cross-request experience
    memories, rather than to a stronger generative backbone.

    \item \textbf{Generative retrieval methods generally outperform
    conventional recommenders.}
    TIGER, FORGE, RankGR, and LoopMemGR achieve substantially higher
    HR than conventional sequential models, particularly at larger
    cutoffs. This pattern is consistent with the advantage of
    generating structured semantic identifiers, which incorporate
    item-level semantic or collaborative information into the
    retrieval process.
\end{itemize}

\subsection{Closed-Loop Experience Memory (RQ2)}
\label{sec:closed_loop_experience}

We examined whether the recommendation experience accumulated from
previous requests benefits subsequent recommendation, and whether TVMR
retains this benefit under a fixed 16-token budget. \emph{Behavior
only} removes the experience log and keeps the behavior sequence
unchanged. \emph{Raw Experience} serializes up to 1,000 experience
entries into the context without any summarization. The remaining
variants read the same experience log and output 16 experience tokens.
They include two generic fixed-budget readers (BlockMean-16 and
AttnPool-16), the long-sequence reader LONGER-16~\cite{chai2025longer},
and view-specific variants of TVMR. All variants share the same
behavior input, RankGR backbone, and evaluation protocol.

\begin{table}[t]
\caption{
Effect of closed-loop experience memory on the \emph{Taobao} dataset.
Raw Experience serializes up to 1,000 experience entries, whereas all
other experience readers output 16 tokens. All values are percentages,
and the best fixed-budget result is boldfaced.
}
\label{tab:experience_memory}
\centering
\resizebox{\columnwidth}{!}{%
\begin{tabular}{lcccc}
\toprule
& \multicolumn{2}{c}{Click HR@$K$}
& \multicolumn{2}{c}{PV HR@$K$} \\
\cmidrule(lr){2-3}
\cmidrule(lr){4-5}
Variant
& $K{=}20$
& $K{=}1000$
& $K{=}20$
& $K{=}1000$ \\
\midrule
Behavior only
& 11.64
& 54.00
& 2.99
& 38.18 \\
\midrule
BlockMean-16
& 12.88
& 57.56
& 3.58
& 41.99 \\

AttnPool-16
& 16.26
& 63.92
& 5.19
& 48.77 \\

LONGER-16
& 20.01
& 67.94
& 6.96
& 53.06 \\
\midrule
Global-16
& 17.29
& 65.51
& 5.68
& 50.47 \\

Recency+Global-16
& 20.07
& 68.22
& 6.99
& 53.37 \\

\textbf{TVMR-16}
& \textbf{20.43}
& \textbf{68.41}
& \textbf{7.16}
& \textbf{53.56} \\
\midrule
Raw Experience (1,000 entries)
& 21.31
& 71.20
& 7.72
& 58.89 \\
\bottomrule
\end{tabular}%
}
\end{table}

Table~\ref{tab:experience_memory} supports four observations:
\begin{itemize}[leftmargin=*, itemsep=2pt, topsep=3pt]
    \item \textbf{Experience memory consistently improves
    recommendation performance.}
    Every experience-augmented variant outperforms \emph{Behavior
    only}, and even simple mean pooling brings clear gains. This
    result confirms that the experience log carries evidence absent
    from the behavior log, which is exactly the one-sided memory gap
    that motivates LoopMemGR.

    \item \textbf{TVMR preserves most of the benefit of raw
    experience within a compact token budget.}
    \emph{Raw Experience} performs best but spends up to 2,000
    additional SID tokens on the serialized log. TVMR retains at least
    74.3\% of its gain over \emph{Behavior only} across the four
    metrics while using only 16 tokens. A compact experience memory
    therefore preserves most of the reusable decision evidence.

    \item \textbf{TVMR is more effective than generic fixed-budget
    readers.}
    TVMR-16 outperforms all alternative readers under the same
    16-token budget, including the strong long-sequence reader
    LONGER-16.This result suggests that organizing experience through task-specific
views is more effective than applying generic compression under the
same token budget.

    \item \textbf{The three memory views provide complementary
    evidence.}
    Anchoring on recent experience yields the largest single
    improvement over the global view alone, and adding the frequency
    view provides further consistent gains on all four metrics.
\end{itemize}

\subsection{In-depth Analysis (RQ3 \& RQ4)}
\label{sec:indepth_analysis}

To further interpret how TVMR retrieves and summarizes
request-relevant evidence from the experience log, we analyze its
attention distributions on evaluated requests. We conduct two
analyses. The first examines whether the global queries learn a
division of labor under the diversity regularizers. The second
examines whether the three views read the same log in complementary
patterns.

\textbf{Query Division in the Global View (RQ3).}
\begin{figure}[t]
\centering
\includegraphics[width=\linewidth]{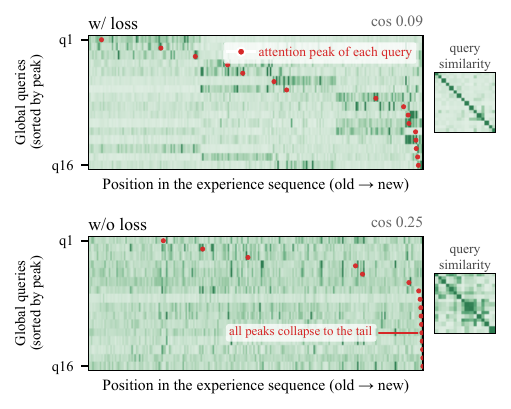}
\caption{Attention of the 16 global queries over the same experience
sequence, trained with (top) and without (bottom) the diversity
regularizers $\mathcal{L}_Q$ and $\mathcal{L}_A$. Rows are sorted by
their peak positions and normalized per row.}
\label{fig:global_loss}
\end{figure}
We visualize the attention of the 16 global queries over the
experience sequence of a request. Rows are sorted
by their peak positions. We also report the mean pairwise cosine
between the 16 attention distributions, where a lower value indicates
less overlapping reading. Table~\ref{tab:diversity_loss} further
compares the retrieval quality of the Global-16 variant trained with
and without the regularizers.

\begin{table}[t]
\caption{Effect of the diversity regularizers on the Global-16
variant. All values are percentages.}
\label{tab:diversity_loss}
\centering
\resizebox{\columnwidth}{!}{%
\begin{tabular}{lcccc}
\toprule
& \multicolumn{2}{c}{Click HR@$K$}
& \multicolumn{2}{c}{PV HR@$K$} \\
\cmidrule(lr){2-3}
\cmidrule(lr){4-5}
Variant
& $K{=}20$
& $K{=}1000$
& $K{=}20$
& $K{=}1000$ \\
\midrule
Global-16 w/o $\mathcal{L}_Q,\mathcal{L}_A$
& 14.31
& 60.18
& 4.26
& 44.78 \\

\textbf{Global-16}
& \textbf{17.29}
& \textbf{65.51}
& \textbf{5.68}
& \textbf{50.47} \\
\bottomrule
\end{tabular}%
}
\end{table}

The result is shown in Figure~\ref{fig:global_loss}, from which we
draw two observations:
\begin{itemize}[leftmargin=*, itemsep=2pt, topsep=3pt]
    \item \textbf{With the regularizers, the global queries partition
    the log into distinct territories.} The sorted rows form a clear
    staircase, and the mean pairwise cosine is 0.09. This pattern is
    stable: across all evaluated requests, the mean pairwise cosine is
    0.13 and never exceeds 0.45. The 16 experience tokens therefore
    summarize different parts of the accumulated experience.

    \item \textbf{Without the regularizers, the queries lose their
    territorial structure.} The pairwise cosine rises to 0.25, and all
    rows read the sequence in a highly overlapping manner. The
    resulting tokens carry redundant evidence, which explains the
    performance drop in Table~\ref{tab:diversity_loss}: Click HR@20
    decreases from 17.29 to 14.31, and PV HR@20 from 5.68 to 4.26,
    with consistent drops at $K{=}1000$.
\end{itemize}

\textbf{View-Level Reading Patterns (RQ4).}
\begin{figure}[t]
\centering
\includegraphics[width=\linewidth]{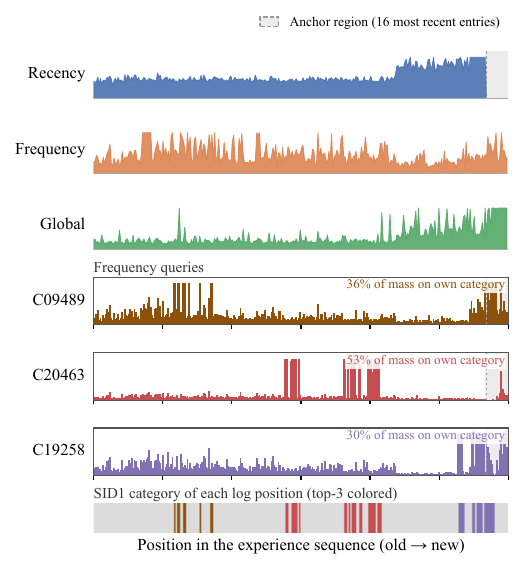}
\caption{How the three views read one experience sequence. Top: the recency, frequency, and global views exhibit complementary attention patterns; the shaded region marks the anchor, i.e., the 16 most
recent entries that are retained directly. Middle: the three largest
frequency queries, each annotated with the share of attention mass
placed on its own category. Bottom: the SID1 category of each log
position, with the three query categories colored.}
\label{fig:case_tracks}
\end{figure}
We further analyze how the three views distribute attention over the
same experience sequence. Figure~\ref{fig:case_tracks} visualizes a request whose experience covers several recurring
categories. The three frequency-query tracks are ordered by the era
of their categories, from the oldest to the newest.

The visualization supports three observations:
\begin{itemize}[leftmargin=*, itemsep=2pt, topsep=3pt]
    \item \textbf{The three views read the log in complementary
    patterns.}
    The recency view covers the whole sequence and assigns greater
    weight to recent entries, while the latest entries are retained
    directly as anchors. The frequency view concentrates on recurring
    categories, whereas the global view selectively attends to a small
    set of entries. Thus, the three views capture different aspects of
    the same experience log.

    \item \textbf{The category prior guides but does not constrain
    frequency retrieval.}
    The three frequency queries place 36\%, 53\%, and 30\% of their
    attention mass on their respective categories, although these
    categories account for only 3\%--8\% of the log. Each query
    nevertheless assigns non-trivial attention outside its own
    category. In particular, the earliest query revisits an old
    interest burst that is no longer emphasized by the recency view.
    The soft category bias therefore supports category-specific
    retrieval while retaining a channel for cross-category evidence,
    consistent with its design.

    \item \textbf{The global view favors recent experience while
    retaining access to informative historical evidence.}
    The global queries assign more attention to recent experience
    overall, but their distributions also exhibit distinct peaks at
    earlier positions in the log. Because these queries are shared
    model-wide rather than constructed for individual requests, this
    pattern suggests that broadly reusable experience is more often
    found in recent interactions, while selected historical entries
    can remain sufficiently informative to attract focused global
    attention.
\end{itemize}

\section{Related Work}
We review two lines of research most relevant to this work: generative recommendation and memory-augmented recommendation.

\subsection{Generative Recommendation}

Generative recommendation formulates item recommendation as the autoregressive generation of discrete item tokens rather than candidate-wise scoring~\cite{geng2022p5,Rajput2023TIGER,Wang2024EAGER,Zhai2024HSTU}. A typical framework first maps items into Semantic IDs (SIDs), represents user histories as token sequences, and then generates the target SID with a sequence model~\cite{Rajput2023TIGER,Wang2024LETTER}. This formulation integrates item representation, user modeling, and candidate retrieval within a unified generative process.

Existing studies have mainly investigated item tokenization and generative modeling. Prior work constructs item tokens through vector quantization, hierarchical indexing, contrastive quantization, textual identifiers, or end-to-end tokenizer learning~\cite{hou2023vqrec,hua2023index,zhu2024cost,tan2024idgenrec,Wang2024LETTER,liu2025etegrec}. Recent methods further incorporate contextual action patterns, heterogeneous behaviors, and large-scale sequence modeling~\cite{hou2025actionpiece,liu2024mbgen,Zhai2024HSTU}. These methods improve how historical behaviors are encoded and generated, but they still reconstruct user state primarily from behavior history at each request. \textsc{LoopMemGR} instead retains recommendation--feedback experience as persistent evidence for subsequent generation.

\subsection{Memory-Augmented Recommendation}

Memory-augmented recommenders preserve long-term user interests through hierarchical storage, recurrent memory, behavior retrieval, or compact interest representations~\cite{Ren2019HPMN,pi2019mimn,Pi2020SIM,Chang2023TWIN}. Recent industrial systems extend this line with retrieval-enhanced comprehension, hierarchical compression, and scalable long-sequence modeling~\cite{Lin2024ReLLa,si2024twinv2,chai2025longer}. Their memory modules mainly summarize or retrieve user-side behavioral evidence, reducing the cost of repeatedly processing an ever-growing history.

LLM-based agents provide a complementary direction by storing explicit memories and using reflection to convert feedback into reusable experience~\cite{shinn2023reflexion,zhang2025memorysurvey}. In recommendation, conversational and agentic methods maintain user profiles, reflect on recommendation errors, coordinate specialized reflectors, or model iterative user feedback~\cite{xi2024memocrs,huang2025interecagent,wang2025re2llm,qin2025more,gu2025r4ec,cai2025feedback}. Interactive recommenders also optimize multi-step trajectories or exploit exposure feedback~\cite{ie2019slateq,gao2023cirs,Zhao2025CaseRec}. However, these approaches either rely on additional LLM reasoning or treat recommendation outcomes as transient signals. \textsc{LoopMemGR} directly maintains system-side recommendation experience across requests and compresses it into bounded recency, frequency, and global memories for an SID-based generative backbone.

\section{Conclusion}
In this paper, we present \textsc{LoopMemGR}, a closed-loop memory framework for generative recommendation that jointly models compact behavioral intent and reusable recommendation experience. Through a \emph{read--recommend--reflect--write} loop, \textsc{LoopMemGR} preserves system-side recommendation decisions and their observed outcomes, enabling subsequent requests to reuse previously accumulated experience under a bounded memory budget. Its multi-view memory reader further captures complementary evidence from recent interactions, recurring recommendation patterns, and transferable regularities shared across users. Extensive experiments on industrial demonstrate the effectiveness of the proposed framework, while ablation studies verify the complementary contributions of intent memory and experience memory. These findings highlight the value of transforming transient recommendation decisions into persistent user-specific experience and provide a promising direction for more stateful generative recommender systems.

\begin{acks}
\end{acks}

\bibliographystyle{ACM-Reference-Format}
\bibliography{8_ref}

\appendix
\section{Datasets.}
\label{dataset}
\begin{table}[t]
    \centering
    \caption{Statistics of the industrial \emph{Taobao} dataset and the public \emph{Beauty} benchmark.}
    \label{tab:dataset_statistics}
    \begingroup
    \setlength{\tabcolsep}{3.8pt}
    \renewcommand{\arraystretch}{1.08}
    \resizebox{\columnwidth}{!}{%
    \begin{tabular}{lrrrr}
        \toprule
        Dataset
        & \#Users
        & \#Items
        & \#Interactions
        & Sparsity \\
        \midrule
        \emph{Taobao}
        & 21 million
        & 270 million
        & 26 billion
        & 99.99\% \\
        \emph{Beauty}
        & 22,363
        & 12,101
        & 198,502
        & 99.93\% \\
        \bottomrule
    \end{tabular}%
    }
    \endgroup
\end{table}

We evaluate LoopMemGR on an industrial dataset, denoted as
\emph{Taobao}, constructed from real-world traffic logs collected from
Taobao's production recommendation system. It contains large-scale user
interactions under multiple feedback signals, including page view,
click, and purchase.
We follow the data construction and partition protocol of
RankGR~\cite{fu2026rankgr}, without introducing additional filtering
or resampling.

To further validate LoopMemGR beyond the industrial setting, we also
evaluate on \emph{Beauty}, a widely used public benchmark from the
Amazon Review dataset\footnote{\url{https://nijianmo.github.io/amazon/index.html}}.
Following common practice, we apply 5-core filtering and treat each
review as an interaction. We adopt the standard leave-one-out protocol:
the last item of each user sequence is held out for testing, the
second-to-last for validation, and the rest for training. Since public
benchmarks provide no exposure logs, the experience memory on
\emph{Beauty} is constructed via a cross-fold exposure replay protocol,
detailed in Appendix~\ref{app:public_replay}.

All interactions are processed in chronological order. For a request
at step $t$, the model can access only the behavioral interactions and
experience entries produced before that request. The target interaction
and the experience generated from the current request are excluded from
its input. The newly written experience becomes available only from
step $t+1$. This chronological protocol prevents future information
from leaking into either the behavioral intent memory or the experience
memory. Detailed dataset statistics are reported in
Table~\ref{tab:dataset_statistics}.

\section{Baselines}
\label{baselines}
We compare LoopMemGR with two categories of recommendation methods:
conventional retrieval methods and generative retrieval methods.

\noindent\textit{(a) Conventional Recommendation Methods.}
\par\nopagebreak

\begin{itemize}[
    labelindent=0.4em,
    leftmargin=*,
    labelsep=0.5em,
    itemsep=1pt,
    topsep=2pt,
    parsep=0pt,
    partopsep=0pt
]
    \item \textbf{YouTubeDNN}~\cite{YouTubeDNN}
    represents a user's preference by aggregating historical item
    embeddings into a dense user vector for efficient large-scale
    candidate retrieval.

    \item \textbf{SASRec}~\cite{Kang2018SASRec}
    employs causal self-attention to model sequential dependencies
    among historical user interactions.

    \item \textbf{BERT4Rec}~\cite{BERT4Rec}
    introduces bidirectional sequence modeling and a masked-item
    prediction objective for sequential recommendation.

    \item \textbf{Caser}~\cite{Caser}
    applies horizontal and vertical convolutional filters over an
    interaction sequence to capture local transitions and
    user-level preference patterns.

    \item \textbf{NextItNet}~\cite{NextItNet}
    uses dilated convolutional networks to efficiently model
    long-range dependencies in sequential behaviors.

    \item \textbf{CORE}~\cite{CORE}
    constructs session representations through weighted aggregation
    and aligns user and item representations in a consistent latent
    space.
\end{itemize}

\noindent\textit{(b) Generative Recommendation Methods.}
\par\nopagebreak

\begin{itemize}[
    labelindent=0.4em,
    leftmargin=*,
    labelsep=0.5em,
    itemsep=1pt,
    topsep=2pt,
    parsep=0pt,
    partopsep=0pt
]
    \item \textbf{HSTU}~\cite{Zhai2024HSTU}
    formulates recommendation as sequential transduction and employs
    hierarchical units to model large-scale behavioral sequences.

    \item \textbf{TIGER}~\cite{Rajput2023TIGER}
    represents each item with a multi-token semantic identifier and
    autoregressively generates the identifier of the next item.

    \item \textbf{FORGE}~\cite{FORGE}
    improves semantic identifier construction by incorporating
    collaborative signals and mitigating identifier collisions in
    industrial item corpora.

    \item \textbf{RankGR}~\cite{fu2026rankgr}
    strengthens generative retrieval through listwise preference
    optimization and a lightweight candidate-aware scoring module
    that refines the initially generated candidates.
\end{itemize}

LoopMemGR is instantiated on top of RankGR. RankGR therefore serves as
the principal backbone-matched baseline: both methods use the same
semantic identifiers, generative backbone, training objective, and
decoding pipeline, while LoopMemGR additionally maintains behavioral
intent memory and cross-request experience memory.

\section{Cross-Fold Exposure Replay on Public Data}
\label{app:public_replay}

Public Amazon-style datasets record only the items that users interacted with, but do not provide system-side exposure logs indicating which items were shown to users. Since the experience memory in LoopMemGR relies on such exposure information, we reconstruct the missing exposure logs by replaying a recommender over each user's interaction history.

The replay procedure consists of three steps. First, we split all users into two folds according to the parity of their user indices and train one simulator for each fold. Each simulator shares the same architecture as our generative backbone. A simulator is trained only on users from its own fold, but is used exclusively to generate exposures for users in the opposite fold. This cross-fold design ensures that the simulator never observes the training targets of the users for whom it generates exposures, thereby preventing label leakage.

Second, we replay each user's interaction history request by request. At each request, the simulator trained on the opposite fold receives only the interactions preceding that request and generates an exposure list using constrained beam search. Feedback is then derived from the ground-truth interaction sequence: an exposed item is labeled \emph{clicked} if it matches the user's next interaction, and \emph{skipped} otherwise. This procedure produces, for every request, an exposure log with corresponding click feedback in the same format as the logs available in the industrial setting.

Third, the reconstructed exposure logs are consumed in exactly the same manner as in the industrial setting. Exposure entries from preceding requests form the experience sequence, which is compressed by TVMR into 16 experience tokens. Each token is augmented with a learnable feedback embedding. To preserve causality, the experience written at a given request is visible only to subsequent requests, following the chronological protocol described in Appendix~\ref{dataset}.

\end{document}